# An electromagnetic-thermal-mechanical coupling model of dry-wound HTS coil based on T-A formulation with Neumann boundary condition

*(This article contains only the theoretical model.)*

Yunkai Tang[1,2,3], Sijian Wang[1,2,3], Donghui Liu[1,2,3], Huadong Yong[1,2,3] and Youhe Zhou[1,2,3]

*1. Key Laboratory of Mechanics on Disaster and Environment in Western China, Ministry of Education of China, Lanzhou University, Lanzhou, Gansu 730000, People's Republic of China*

*2. Department of Mechanics and Engineering Sciences, College of Civil Engineering and Mechanics, Lanzhou University, Lanzhou, Gansu 730000, People's Republic of China*

*3. Institute of Superconductor Mechanics, Lanzhou University, Lanzhou 730000, People's Republic of China*

**Abstract**

The multi-physics coupling behaviours of HTS coils have now received much attention. In particular, the electromagnetic field, temperature field and mechanical deformation interact with each other during quench of high-field magnets. Accurate analysis of coupling behaviours becomes the key to designing magnets and quench protection. In this paper, a multi-physics coupling model is proposed based on T-A formulation with Neumann boundary conditions. It is convenient to analyse the effects of deformation and temperature on the electromagnetic field, as well as the redistribution of the current between the different layers during quench.

## 1. Introduction

The multi-physics coupling behaviours of HTS magnets have now received much attention. In high magnetic fields, the large Lorentz force will cause the deformation of superconducting magnets, which will lead to the change of electromagnetic field distribution and the degradation of critical current. The electromagnetic-mechanical model based on the T-A formulation has been developed, which can characterize experiments effectively [1-4]. However, heat is generated during quench, thus change of temperature needs to be considered in the simulations. Since the redistribution of current between different layers is induced by quench, it is necessary to modify the T-A formulation. The electromagnetic-thermo-mechanical coupling model can also be established based on the modified T-A formulation.

Recently, a modified T-A formulation was proposed with the Neumann boundary condition [5], which can conveniently calculate the current sharing in the tape. In the process of quench, the current sharing can also be dealt with the modified T-A formulation. In this paper, an electromagnetic-thermo-mechanical coupling model is presented with the modified T-A formulation to analyse the multi-physics coupling behaviours of coil during quench.

## 2. Theory

*2.1. T-A formulation with Neumann boundary condition*

T-A formulation can calculate the electromagnetic field of ReBCO coated conductor efficiently, and the government equations are

$$\nabla \times (\rho \nabla \times \mathbf{T}) = -\frac{\partial \mathbf{B}}{\partial t}, \quad \nabla^2 \mathbf{A} = -\mu \mathbf{J}, \quad \mathbf{B} = \nabla \times \mathbf{A}, \quad \mathbf{J} = \nabla \times \mathbf{T}, \quad (1)$$

where $\mathbf{J}$, $\mathbf{T}$, $\mathbf{B}$ and $\mathbf{A}$ are respectively the current density, the electric vector potential, the magnetic induction intensity, and the magnetic vector potential. $\mu$ is the magnetic permittivity and $\rho$ is the electrical resistivity of the individual components of the ReBCO tape. Note that for YBCO, the letter is

$$\rho = \begin{cases} \dfrac{\rho_{sc} \cdot \rho_n}{\rho_{sc} + \rho_n} & T < T_c \\ \rho_n & T \geq T_c \end{cases}, \qquad \rho_{sc} = \dfrac{E_c}{J_c}\left(\dfrac{|\mathbf{J}|}{J_c}\right)^{n-1}, \qquad (2)$$

where $T$ is the temperature, $E_c = 1\ \mu\text{V/cm}$, the critical temperature $T_c = 92\ \text{K}$ and the critical current density is

$$J_c = J_{c0} \cdot J_c(\mathbf{B}) \cdot J_c(\varepsilon) \cdot J_c(T), \qquad (3)$$

where $J_{c0}$ and $\varepsilon$ are respectively the initial critical current density the strain along the length direction of the tape. Author components of $J_c$ are

$$\begin{aligned} J_c(\mathbf{B}) &= \dfrac{1}{\left[1 + \sqrt{(aB_\parallel)^2 + B_\perp^2}\Big/B_c\right]^b} \\ J_c(\varepsilon_\theta) &= \begin{cases} 1 - 6.28\times 10^{-8}\exp(2174.35\varepsilon_\theta) & \varepsilon_\theta \leq 0.67\% \\ 0.867 - 582.466\times(\varepsilon_\theta - 0.0067) & \varepsilon_\theta > 0.67\% \end{cases}, \\ J_c(T) &= \begin{cases} \dfrac{T_c - T}{T_c - T_0} & T \leq T_c \\ 0 & T > T_c \end{cases} \end{aligned} \qquad (4)$$

where $B_\parallel$ and $B_\perp$ are respectively the parallel and perpendicular components of $\mathbf{B}$. The value of parameters $n$, $a$, $b$, $B_c$, $T_0$ and $J_{c0}$ in Eqs. (2) and (4) depend on the tape and the environment, as shown in the examples in the following text.

Eq. (1) is written in the axisymmetric form due to the shape of the pancake coil. Because of the thin structure of the ReBCO tape, the electromagnetic induction caused by $B_\parallel$ is almost negligible, and only $B_\perp$ appears in Eq. (1), which changes with the deformation of the tape, as shown in Fig. 1. In addition, the parameter $a$ in $J_c(\mathbf{B})$ is smaller than 1, which means the electromagnetic constitution of ReBCO is sensitive to $B_\perp$. Therefore, it is essential to evaluate $B_\perp$ accurately. The electromagnetic-mechanical coupling model proposed by Kolb-Bond and Yan [ref] defined the deflection angle $\beta = \partial u_r/\partial z$ of the tape, where $u_r$ is the radial displacement. $B_\perp$ can be obtain by the radial field $B_r$ and the axial field $B_z$ as follows:

$$B_\perp = B_r \cos\beta - B_z \sin\beta . \tag{5}$$

Usually, the current is in YBCO due to its tiny electrical resistivity. However, other components will also carry current during quench, and the layers of ReBCO are parallel. It makes the above T-A formulation not computable directly. Fortunately, the calculation of shunt behaviors is solved in our previous work [5], which proposed a modified formulation with the Neumann boundary condition. Its weak form is

$$\int_\Omega (\nabla \times \delta\mathbf{T}) \cdot \rho(\nabla \times \mathbf{T}) \mathrm{d}\Omega + \int_\Omega \frac{\partial \mathbf{B}}{\partial t} \cdot \delta\mathbf{T} \mathrm{d}\Omega + \int_\Gamma (\mathbf{E} \times \delta\mathbf{T}) \cdot \mathrm{d}\mathbf{\Gamma} = 0, \tag{6}$$

where $\Omega$ and $\Gamma$ are respectively the domain and the boundary. When the cross area of each layer is modeled as line segments, $\Gamma$ represents the endpoint. $\int_\Gamma (\mathbf{E} \times \delta\mathbf{T}) \cdot \mathrm{d}\mathbf{\Gamma}$ is the Neumann boundary term. Integral constraints are added to replace the Dirichlet boundary conditions in the original T-A formulation to ensure that the total current in each turn is equal to $I_{op}$. Other details are provided in [5]. In addition, the coupling ways of the original T-A formulation are also applicable here.

*2.2. Contact mechanics model*

For the dry-wound coil, the contact behaviours between turns will affect the deformation of the tape. Consider any two bodies in contact, denoted "*Body 1*" and "*Body 2*"; they mean the coil's tape and insulation respectively. Since the electromagnetic loading of the ReBCO coil is slow, the inertia term is not considered, the weak form (i.e., the virtual work principle) based on the finite deformation theory in the reference configuration is

$$\int_{\Omega_1 \cup \Omega_2} \boldsymbol{\tau} : \delta\boldsymbol{\varepsilon} \mathrm{d}\Omega - \int_{\Omega_1 \cup \Omega_2} \mathbf{f}^L \cdot \delta\mathbf{u} \mathrm{d}\Omega - \int_{\Gamma_1^f \cup \Gamma_2^f} \mathbf{f}^b \cdot \delta\mathbf{u} \mathrm{d}\Gamma - \int_{\Gamma_1^c \cup \Gamma_2^c} \mathbf{f}^c \cdot \delta\mathbf{u} \mathrm{d}\Gamma = 0, \tag{7}$$

where $\boldsymbol{\tau}$, $\boldsymbol{\varepsilon}$ and $\mathbf{u}$ are respectively the second Piola-Kirchhoff (PK-II) stress, the Green-Lagrange (GL) strain and the displacement. $\mathbf{f}^L = \mathbf{J} \times \mathbf{B}$ is the Lorentz force, which is obtained from T-A formulation. $\mathbf{f}^b$ and $\mathbf{f}^c$ are respectively the known boundary force and the unknown contact force. $\Gamma^f$ and $\Gamma^c$ are the corresponding surfaces on Body 1 and 2. In the first term on the left of Eq. (7), the GL strain $\boldsymbol{\varepsilon}$ and the PK-II stress $\boldsymbol{\tau}$ satisfy the constitutive relation

$$\boldsymbol{\varepsilon} = \boldsymbol{\varepsilon}_{el} + \boldsymbol{\varepsilon}_{th} = \mathbf{S} : \boldsymbol{\tau} + \int_{T_0}^{T} \alpha(T) \mathrm{d}T, \tag{8}$$

where $\mathbf{S}$ is the flexibility tensor and $\alpha$ is the thermal expansion coefficient changing with temperature. If the small deformation theory is considered, the GL strain $\boldsymbol{\varepsilon}$ becomes

$$\boldsymbol{\varepsilon} = \frac{1}{2}(\mathbf{u} \otimes \nabla + \nabla \otimes \mathbf{u} + \nabla \otimes \mathbf{u} \cdot \mathbf{u} \otimes \nabla) \approx \frac{1}{2}(\mathbf{u} \otimes \nabla + \nabla \otimes \mathbf{u}), \tag{9}$$

which can improve the efficiency of numerical calculation.

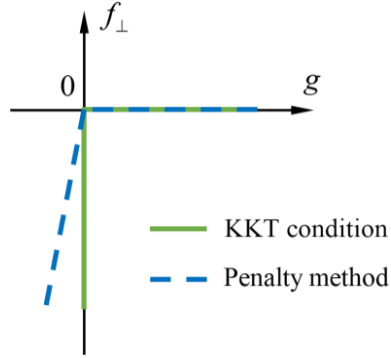

**Figure 2.** The relationship between $f_\perp$ and $g$.

The last term on the left of Eq. (7) can be expressed as

$$-\int_{\Gamma_1^c \cup \Gamma_2^c} \mathbf{f}^c \cdot \delta \mathbf{u} \, \mathrm{d}\Gamma = -\int_{\Gamma_1^c} \mathbf{f}_1^c \cdot \delta \mathbf{u}_1 \, \mathrm{d}\Gamma - \int_{\Gamma_2^c} \mathbf{f}_2^c \cdot \delta \mathbf{u}_2 \, \mathrm{d}\Gamma = \int_{\Gamma_1^c} \mathbf{f}_1^c \cdot (\delta \mathbf{u}_2 - \delta \mathbf{u}_1) \mathrm{d}\Gamma. \tag{10}$$

Note that Newton's Third Law is considered. If the friction is ignored, Eq. (10) is

$$\int_{\Gamma_1^c} f_\perp \mathbf{n}_1 \cdot (\delta \mathbf{u}_2 - \delta \mathbf{u}_1) \mathrm{d}\Gamma = \int_{\Gamma_1^c} f_\perp \cdot \delta g \, \mathrm{d}\Gamma, \tag{11}$$

where $\mathbf{n}_1$ is the normal vector of $\Gamma_1^c$ and $f_\perp$ is the component of $\mathbf{f}_1^c$ along $\mathbf{n}_1$. The gap between the tape and the insulation is

$$g = \mathbf{n}_1 \cdot (\mathbf{x}_2 - \mathbf{x}_1) = \mathbf{n}_1 \cdot [(\mathbf{X}_2 + \mathbf{u}_2) - (\mathbf{X}_1 + \mathbf{u}_1)], \tag{12}$$

where $\mathbf{x}$ and $\mathbf{X}$ are the position vectors in the current configuration and the reference configuration of the material point. Note that $\delta \mathbf{X}$ is equal to zero. In addition to the virtual work equation, the following equations, which are also called Karush-Kuhn-Tucker (KKT) conditions and have intuitive physical meanings, are to be supplemented,

$$g \geq 0, \quad f_\perp \leq 0, \quad g \cdot f_\perp = 0. \tag{13}$$

Therefore, the relationship between $f_\perp$ and $g$ is shown as the solid line in Fig. 2. It can be seen that $f_\perp$ is a multi-valued function when $g = 0$. To calculate Eq. (13) numerically, a single-valued function is defined as

$$f_\perp = \begin{cases} \kappa g & g < 0 \\ 0 & g \geq 0 \end{cases}, \quad \kappa > 0. \tag{14}$$

As shown in the dotted line in Fig. 2, the function approximates the KKT condition when the penalty factor $\kappa$ is large, and the impenetrable condition (i.e., $g \geq 0$) is violated slightly.

*2.3. Solid heat transfer model*

The temperature $T$ is calculated by the following equation

$$\gamma C(T) \cdot \frac{\partial T}{\partial t} - \nabla \cdot \left[ \mathbf{K}(T) \cdot \nabla T \right] = \mathbf{E} \cdot \mathbf{J} + Q, \tag{15}$$

which is the axisymmetric form for the pancake coil. In Eq. (15), $\gamma$ is the mass density, $C$ and $\mathbf{K}$ are respectively the heat capacity and the thermal conductivity changing with temperature. $Q$ is the pulsed energy generated by the heater to induce the quench.

*2.4. Homogenization technique*

The homogenization technique is used because the multi-layer structure of ReBCO makes the complete coil model calculation very time-consuming. As shown in Fig. 3, the layers on either side of YBCO are homogenized as composite materials. The equivalent resistivity of the composite material can be calculated from a parallel circuit

$$\rho = \left( \sum_{i=1}^{m} \frac{d_i}{d} \cdot \frac{1}{\rho_i} \right)^{-1}, \qquad (16)$$

where $m$ is the number of layers in the composite material, and $d$ is the thickness of the composite material. $\rho_i$ and $d_i$ are respectively the electrical resistivity and the thickness of the $i$-th layer. The equivalent mechanical parameters of the two composites can be obtained by applying unit strain to representative volume element (RVE), and the first term of the constitutive equation Eq. (8) becomes

$$\begin{pmatrix} \varepsilon_r \\ \varepsilon_\theta \\ \varepsilon_z \\ \varepsilon_{rz} \end{pmatrix} = \begin{pmatrix} \frac{1}{Y_r} & -\frac{\upsilon_{r\theta}}{Y_\theta} & -\frac{\upsilon_{rz}}{Y_z} & 0 \\ -\frac{\upsilon_{\theta r}}{Y_r} & \frac{1}{Y_\theta} & -\frac{\upsilon_{\theta z}}{Y_z} & 0 \\ -\frac{\upsilon_{zr}}{Y_r} & -\frac{\upsilon_{z\theta}}{Y_\theta} & \frac{1}{Y_z} & 0 \\ 0 & 0 & 0 & \frac{1}{G_{rz}} \end{pmatrix} \begin{pmatrix} \tau_r \\ \tau_\theta \\ \tau_z \\ \tau_{rz} \end{pmatrix}, \qquad (17)$$

where $Y$, $G$ and $\upsilon$ and are respectively the equivalent Young's modulus, shear modulus and Poisson's ratio. Anisotropic thermal expansion coefficients are

$$\alpha_\eta = \frac{\sum_{i=1}^{m} d_i E_i \alpha_i}{\sum_{i=1}^{m} d_i E_i} \qquad \alpha_r = \sum_{i=1}^{m} (1+\upsilon_i) \cdot \frac{d_i \alpha_i}{d} - \upsilon_{r\eta} \alpha_\eta \qquad \eta = \theta, z, \qquad (18)$$

where $E_i$, $\upsilon_i$ and $\alpha_i$ are respectively the Young's modulus, Poisson's ratio and thermal expansion coefficient of the $i$-th layer. The equivalent mass density is

$$\gamma = \sum_{i=1}^{m} \frac{d_i \gamma_i}{d}, \qquad (19)$$

where $\gamma_i$ is the mass density of the $i$-th layer. For the heat transfer, the equivalent thermal conductivity $\mathbf{K}$ and the equivalent heat capacity $C$ are respectively

$$\mathbf{K} = \begin{pmatrix} k_r & 0 & 0 \\ 0 & k_\varphi & 0 \\ 0 & 0 & k_z \end{pmatrix} \qquad k_r = \left( \sum_{i=1}^{m} \frac{d_i}{d} \cdot \frac{1}{k_i} \right)^{-1} \qquad k_\varphi = k_z = \sum_{i=1}^{m} \frac{d_i}{d} \cdot k_i, \qquad (20)$$

$$C = \sum_{i=1}^{m} \frac{d_i}{d} \cdot C_i,  \tag{21}$$

where $k_i$ and $C_i$ are respectively the thermal conductivity and the heat capacity of the *i*-th layer in the composite material.

The coupling mode of the above physical fields is shown in Fig. 4.

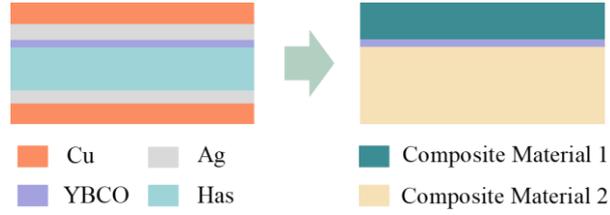

**Figure 3.** The homogenization model of ReBCO tape.

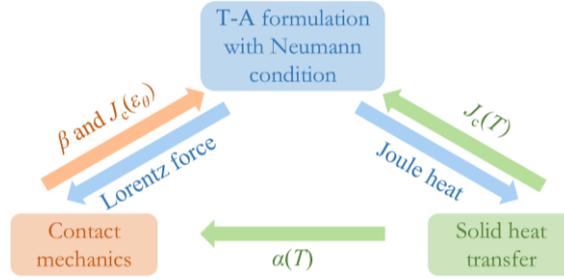

**Figure 4.** Schematic of the coupling model.

[1] Y. Li, D. Park, W. Lee, Y. Choi, H. Tanaka, J. Bascunan, Y. Iwasa, Screening-Current-Induced Strain Gradient on REBCO Conductor: An Experimental and Analytical Study with Small Coils Wound with Monofilament and Striated Multifilament REBCO Tapes, IEEE Transactions on Applied Superconductivity, 30 (2020) 1-5.

[2] D. Kolb-Bond, M. Bird, I.R. Dixon, T. Painter, J. Lu, K.L. Kim, K.M. Kim, R. Walsh, F. Grilli, Screening current rotation effects: SCIF and strain in REBCO magnets, Superconductor Science and Technology, 34 (2021).

[3] Y. Yan, P. Song, C. Xin, M. Guan, Y. Li, H. Liu, T. Qu, Screening-current-induced mechanical strains in REBCO insert coils, Superconductor Science and Technology, 34 (2021).

[4] M. Niu, H. Yong, Y. Zhou, 3D modelling of coupled electromagnetic-mechanical

responses in REBCO coils involving tape inhomogeneity, Superconductor Science and Technology, 35 (2022).

[5] S. Wang, H. Yong, Y. Zhou, Calculations of the AC losses in superconducting cables and coils: Neumann boundary conditions of the T–A formulation, Superconductor Science and Technology, 35 (2022).

[6] H. Zhang, M. Zhang, W. Yuan, An efficient 3D finite element method model based on the T–A formulation for superconducting coated conductors, Superconductor Science and Technology, 30 (2017).

[7] Y. Wang, W.K. Chan, J. Schwartz, Self-protection mechanisms in no-insulation (RE)Ba2Cu3Oxhigh temperature superconductor pancake coils, Superconductor Science and Technology, 29 (2016).

[8] D. Liu, W. Wei, Y. Tang, H. Yong, Y. Zhou, Delamination behaviors of an epoxy-impregnated REBCO pancake coil during a quench, Engineering Fracture Mechanics, 281 (2023).